# IMPLEMENTATION OF A CLOUD DATA SERVER (CDS) FOR PROVIDING SECURE SERVICE IN E-BUSINESS

D.Kesavaraja[1] , R.Balasubramanian[2] and D.Sasireka[3]

[1] Lecturer , Department of Computer Science and Engineering,
Dr Sivanthi Aditanar College of Engineering , Tiruchendur
*dkesavaraja@gmail.com*
[2]Associate Professor, Department of Computer Science and Engineering,
Manonmaniam Sundaranar University, Tirunelveli
*rbalus662002@yahoo.com*
[3]Lecturer , Department of Information Technology
PSN College of Engineering and Technology , Tirunelveli
*edsasireka@yahoo.com*

## ABSTRACT

*Cloud Data Servers is the novel approach for providing secure service to e-business .Millions of users are surfing the Cloud for various purposes, therefore they need highly safe and persistent services. Usually hackers target particular Operating Systems or a Particular Controller. Inspiteof several ongoing researches Conventional Web Servers and its Intrusion Detection System might not be able to detect such attacks. So we implement a Cloud Data Server with Session Controller Architecture using Redundancy and Disconnected Data Access Mechanism. In this paper, we generate the hash code using MD5 algorithm. With the help of which we can circumvent even the attacks, which are undefined by traditional Systems .we implement Cloud Data Sever using Java and Hash Code backup Management using My SQL. Here we Implement AES Algorithm for providing more Security for the hash Code. The CDS using the Virtual Controller controls and monitors the Connections and modifications of the page so as to prevent malicious users from hacking the website. In the proposed approach an activity analyzer takes care of intimating the administrator about possible intrusions and the counter measures required to tackle them. The efficiency ratio of our approach is 98.21% compared with similar approaches.*

## 1. INTRODUCTION

With the tremendous growth of Web-based services and sensitive information on web, Web security is getting more important than ever. Web Based Applications has become essential in everyday life. People use the Cloud to work, to exchange information, to make purchases, etc. This growth of the Cloud use has unfortunately been accompanied by a growth of malicious activity in the Cloud . More and more vulnerabilities are discovered, and nearly every day, new security advisories are published.

Potential attackers are very numerous, even if they represent only a very small proportion among the hundreds of millions of Cloud users and clients.

The problem is thus particularly tricky: on one hand, the development of the Cloud allows complex and sophisticated services to be offered, and on the other hand, these services offer to the attacker many new weaknesses and vulnerabilities to exploit.

The complexity of current computer systems has been causing an immense number of vulnerabilities. The number of cyber-attacks has been growing making computer security as a whole an important research challenge.

## 2. INTRUSION TOLERANCE

Database can be defined as the capability of a system to provide reliable service even after an intrusion attack has taken place or has affected a part of a system.

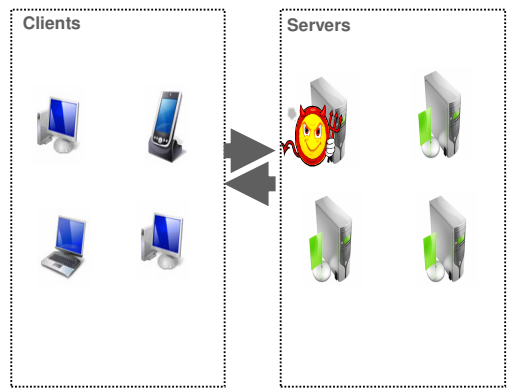

Fig 1 – Intrusion Tolerance

Fig-2 gives the detailed structure intrusion taken place at one server and other servers taking care of service providing.

A dependable system [1] is defined as one that is able to deliver a service that can justifiably be trusted. Attributes of dependability include availability (readiness for correct service), reliability (continuity of correct service), confidentiality (prevention of unauthorized disclosure of information), and integrity (the absence of improper system state alterations). Security is the concurrent existence of

1) Availability

2) Confidentiality

3) Integrity

### 2.1. HTTP Web Server

Distributed Data backups are managed by Web Server and provide a reliable service to the user. Redundancy is used to increase system availability.

Most attacks take advantage of specific vulnerabilities in a particular OS, controller , or hardware platform, they are, in general, ineffective on others. So, the deployment of a redundant data management Web servers (hardware/OS/Virtual Controller ) should allow the system to continue providing acceptable service to users, even if parts of the system are corrupted. The Web servers provide the same services but run different platforms.

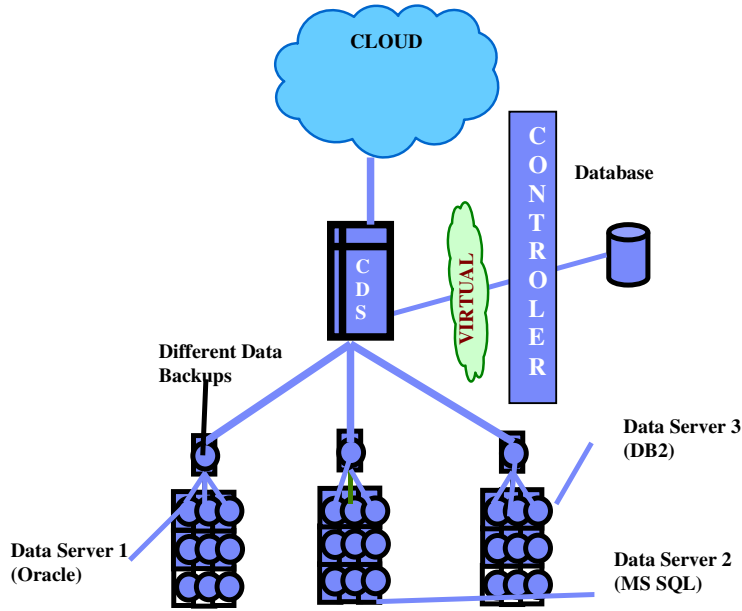

Fig 2 – Distributed Web Server Architecture

Fig-2 gives the detailed structure of CDS and how the virtual controller is connected with the CDS.

## 2.2. Cloud Data Sever

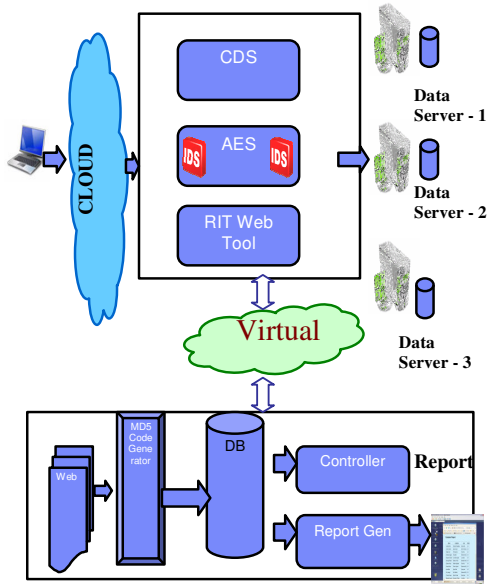

Fig 3 – Cloud Data Server

Fig-3 gives the detailed structure and working of the CDS.

Cloud Data Server provides service taking care of security and persistent availability required for a web service. When the CDS gets a HTTP request, instead of immediately providing the HTTP response it holds the request and connects to the virtual controller .When the virtual controller is connected, the CDS is cut off from the cloud. After the Agreement protocol is satisfied the virtual controller is cut off and the HTTP response is sent to the user from the CDS. Thus the controller were the hash code resides is segregated from the cloud during the agreement protocol process ensuring inability to hack the system.

### 2.2.1. Virtual Controller

The Virtual Controller is used to monitor the web server and analyse its status. If any failure occurs, based on the level of failure it generates an alarm to the administrator .This process should be done through the following algorithm

1. Monitor Web Servers at every Second as well as every Request

2. Store the Data's about the failure

3. If Failure Occurs, automatically it will take remedy based on Agreement Protocol or Change and Response Protocol. It will create an report in Control sender and analyse the failure rate

4. If Failure-Rate < $TH_L$ , Save the report in Cloud Server

5. Else If $TH_L$< Failure-Rate < $TH_U$, Save the report in Cloud Server and Send a beep alert to the Controller.

6. Else If Failure-Rate > $TH_U$ , Save the report in Cloud Server and Send an Higher beep alert to the Controller

### 2.2.2. Hash Code Generator [ Md5 ]

An MD5 hash is typically expressed as a 32 digit Hexadecimal number. MD5 consists of 64 of these operations, grouped in four rounds of 16 operations. F is a nonlinear function; one function is used in each round. Mi denotes a 32-bit block of the message input, and Ki denotes a 32-bit constant, different for each operation.<<<s denotes a left bit rotation by s places; s varies for each operation. Denote addition modulo $2^{32}$.

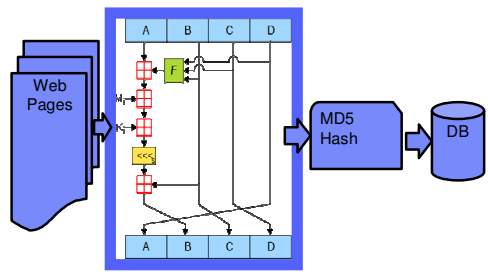

Fig 4 – MD5 Hash Code Generation

Fig-4 describes the implementation of MD5 Algorithm in hash code generation and storage.

### 2.2.3. Agreement Protocol

Agreement protocol ensures that reliable service is provided to the users by Hash code mechanism and virtual controller. It is used to provide a Continual and Safe Service to the user. Its algorithm is as follows:

1. CDS reads the http request

2. Holds the request and disconnects the cloud.

3. Controller is connected with the CDS.

4. CDS checks the Hash Code, Generated by the controller for sub server 1 along with the already stored hash code in the database.

5. If a Match Occurs , Then the controller is disconnected and http response is sent to the user by connecting to the cloud

6. If Match Fails then step 4 is repeated for sub server 2, and the controller starts replacing the affected file with the original file

7. If Match Fails then step 4 is repeated for sub server 3. and the controller starts replacing the affected file with the original file

8. If Match fails then System is rebooted.

**2.2.4. Change And Response Protocol**

The Protocol ensures that the controller checks the web servers periodically or at a specified time based upon traffic flow to ensure that no intrusion has affected the servers .Its algorithm is as follows:

1. When the Traffic flow is very low the cloud is disconnected.

2. Controller is connected with the CDS.

3. CDS checks the Hash Code, Generated by the controller for sub server 1, 2, 3 along with the already stored hash code in the database.

4. If a Match Occurs, Then the controller is disconnected and cloud is activated.

5. If Match Fails then , the controller starts replacing the affected file with the original file

**2.2.5. AES Algorithm**

AES algorithm ensures that the hash code is encrypted in a highly secure manner. AES has a fixed block size of 128 bits and uses a key size of 128 in this paper. Its algorithm is as follows:

1. Key Expansion

2. Initial Round

3. Add Round Key

4. Rounds

5. Sub Bytes—a non-linear substitution step where each byte is replaced with another according to a lookup table.

6. Shift Rows—a transposition step where each row of the state is shifted cyclically a certain number of steps.

7. Mix Columns—a mixing operation which operates on the columns of the state, combining the four bytes in each column

8. Add Round Key—each byte of the state is combined with the round key; each round key is derived from the cipher key using a key schedule.

9. Final Round (no Mix Columns)

10. Sub Bytes

11. Shift Rows

12. Add Round Key

### 2.3. Data Base Manager - DBM

Database Manager Monitors the activity of Virtual Controller and performs the following operations

1. Activity Analysis

2. Alarm Generation

3. Primary Sub Server Selection

4. Pictorial Status Representation

The Intrusion Manager includes the following steps:

A CRP monitor receives an alert "Sub Server S is corrupted" and sends a vote request to the AP module.

2   The vote finishes with a consensus on the corruption of S.

3   The vote result is broadcast to all Sub Servers.

4   The AP module sends a request to the Server manager to generate the list of countermeasures to react to this intrusion.

5   The countermeasures list is broadcast to all Sub Servers.

6   The Virtual Controller sends to the specific module the orders to isolate S and to change the current regime. The Virtual Controller also changes the CRP frequency.

7   The Alarm generator receives the alert notification sent by S after its regime change.

### 2.3.1. Activity Analysis

The DBM records the information for various activities taking place in the CDS and analyses this information to provide an activity analysis .Some of the activities that are monitored are as follows:

HTTP Request :

1. Date of Request
2. Time of Request
3. Content Type
4. Web Page Requested
5. Session Time
6. Status of Request
7. Status of Response
8. Intrusion Activity
9. Infected Sub Server
10. Infected Web Page
11. Frequency of Attacks
12. Reliability of Server

### 2.3.2. Alarm Generation

Based upon the activity analysis the status of alarm is decided and corresponding alarm is aroused.

Example:

If the Failure rate $<TH_L$ higher level alarm is raised to intimate the administrator for necessary actions.

### 2.3.3. Primary Sub Server Selection

The Activity Analysis Process rates the Sub Servers into the following types

Trust worthy – High Priority

Suspected   - Medium Priority

Corrupted   - Low Priority

The Primary Sub Server Selector selects the Primary Sub Server based on the priority and Load balance .The Priority of a server is decided based upon the following formula:

Initial State of Sub Servers

$AS(X) = 0$   - Trust Worthy ----- (1)

When the first Alert occur the system moves to the suspected states

$AS(X) += Awi(x)$ –Suspected ----- (2)

It remains in the suspected state until the following condition satisfies

$AS(X) = Awi(X) > TC$- Corrupted ----- (3)

When the condition fails the system moves to the corrupted state and the system is rebooted after which it moves back to trust worthy state.

AS(X)  :   alert level of Sub Server X

Awi        :   weight of alert i for

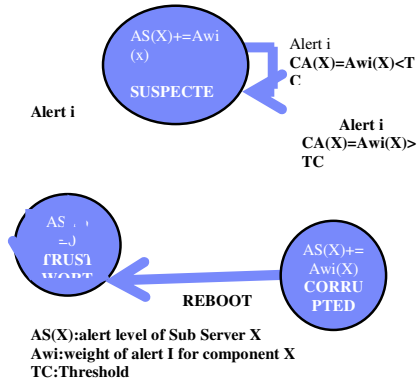

AS(X):alert level of Sub Server X
Awi:weight of alert I for component X
TC:Threshold

Fig 6– State Management

Fig 6 shows the transformation of a sub server to various states.

## 3. IMPLEMENTATION

This CDS is implemented Using Java and My SQL as a backend. It provides a reliable and highly secure Service to the user .The Architecture is distributed thus providing continual Secure Service to all clients.

### 3.1. CDS Server

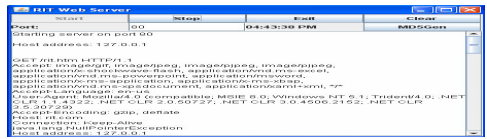

Fig 7 – Snapshot of CDS Server

Fig-7 shows the snapshot of CDS server displaying the information about the request.

### 3.2 CDS Web Service

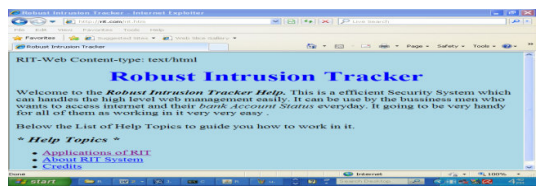

Fig 8 –CDS Web Service

Fig-8 shows a sample webpage provided as a service by the CDS server.

### 3.3. CDS Activity Analysis

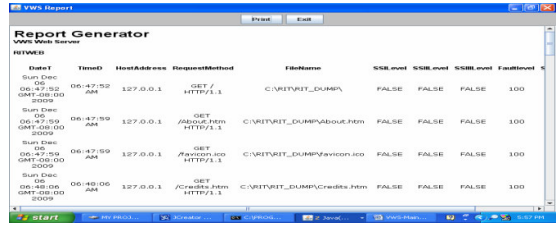

Fig 9 –CDS Activity Analysis

Fig-9 shows the Activity Analysis of the CDS Server on a particular instance. Sample webpage provided as a service by the CDS server.

### 3.4. CDS Alarm Generation

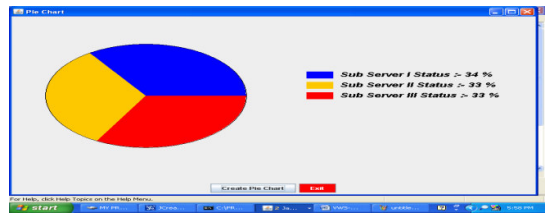

Fig 10 –CDS Alarm Generation

Fig-10 shows an Alarm generated at a particular instance when a server has been infected.

### 3.5. CDS Pictorial Representation

Fig 11 –CDS Pictorial Representation

Fig-11 shows the pictorial representation of the Activity analysis based upon information stored at a point of time.

## 4. PERFORMANCE ANALYSIS

Performance analysis for CDS was performed considering the following four factors.

## 4.1. Time Complexity

TABLE 1

TIME COMPLEXITY ANALYSIS

| Sl No | Name of the Web page | Page Size (MB) | Web Server (ns) | Web Server (ns) | Web Server (ns) |
|---|---|---|---|---|---|
| 1 | http:\\RITWeb .com | 2.14 | 74 | 91 | 81 |
| 2 | http:\\HelpWorks.com | 3.12 | 92 | 112 | 93 |
| 3 | http:\\E-Learn .com | 4.15 | 112 | 123 | 110 |
| 4 | http:\\JavaEducation.com | 5.3 | 120 | 142 | 131 |
| 5 | http:\\webedu .com | 5.7 | 131 | 153 | 142 |

Table 1 shows the time taken for loading the various pages using different web servers.

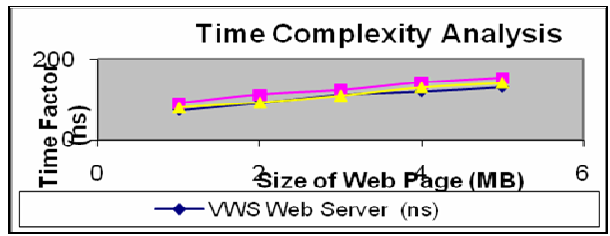

Fig 13 – Time Complexity Analysis

Fig-13 shows the time taken for loading the various pages using different web servers pictorially.

## 4.2. Persistent service Availability

Persistent service Availability lies between 0 to 1. It is been measured using Rank Correlation by the relation between web servers.

TABLE 2

PERSISTENT SERVICE AVAILABILITY

| Sl No | Name of the Web page | Persistent service Availability (Rank Correlation ) | |
|---|---|---|---|
| | | Average Delay Time | Average Speed |
| 1 | Cloud Data Server | 45 | 13 |
| 2 | Http Web Server | 93 | 27 |
| 3 | Java Web Server | 102 | 31 |

Table 2 shows the Persistent service Availability of various servers.

$$r = 1 - \frac{6\sum d^2}{N(N^2-1)} \qquad (4)$$

r-Rank Correlation

d-differences in Rank

N-Number of Servers

Using the formula (4) Rank correlation is calculated. The Value of r=1.

### 4.3. Algorithm Performances

MD5 Algorithm is the one of the powerful algorithm to generate hash code [1].

 AES Algorithm is the one of the most secure encryption algorithm [6].

## 5. APPLICATIONS

It can be applied in areas were continual secure service is required. Example E-Business, Online Shopping and Online Share Trading, Auction and, It can also be applied in areas were continual reliable service is required. Example Cash Transaction, Online Shopping and E-Governance

A recent Survey during November 2009 predicts that around 698 websites have vanished due to improper security features. Using CDS   the security provided increases in large fold.

## 6. FUTURE WORK

In this paper the Emerald [5] tool has been used to stop malicious scripts intruding and causing damage. Usage of Tools like Emerald requires constant up gradation to stop new way of attacks. In the future a mechanism can be developed for stopping the same without External tool.

## 7. CONCLUSION

CDS server was tested against standard webs servers in order to rate it. From the analysis it has been found that CDS Server stands unique in providing safe and persistent service to the user compared to the other web servers. A new mechanism named disconnected data access has been introduced to provide increased security. MD5 and AES algorithms used for creating hash codes and encrypting them is the best algorithm for the Process. The virtual controller increases the reliability for using disconnected method. The Activity Analyser helps the administrator time to time in knowing about the intrusion caused and its counter measures. Our Project is efficient to a mark of 98.21% comparing others.

### REFERENCES

[1] "THE DESIGN OF A GENERIC INTRUSION TOLERANT ARCHITECTURE FOR WEB SERVERS "By Ayda Saidane, Vincent Nicomette, And Yves Deswarte, Member, IEEE, IEEE TRANSACTIONS ON DEPENDABLE AND SECURE COMPUTING, VOL. 6, NO. 1, JANUARY-MARCH 2009

[2] "DATA FUSION AND COST MINIMIZATION FOR INTRUSION DETECTION "By Devi Parikh, Student Member, IEEE, and Tsuhan Chen, Fellow, IEEE, IEEE TRANSACTIONS ON INFORMATION FORENSICS AND SECURITY, VOL. 3, NO. 3, SEPTEMBER 2008

[3] "AN ARCHITECTURE FOR AN ADAPTIVE INTRUSION-TOLERANT SERVER " By Alfonso Valdes, Magnus Almgren, Steven Cheung, Yves Deswarte ,Bruno Dutertre, Joshua Levy, Hassen Sadi, Victoria Stavridou, and Tomas E. Uribe

[4] "GRAPHICAL INFERENCES FOR MULTIPLE INTRUSION DETECTION " By Tung Le , Student Member , IEEE , and Christoforos N.Hadjicostis , Senior Member , IEEE

[5] "RANDOM-FOREST-BASED NETWORK INTRUSION DETECTION SYSTEMS " By Jiong Zhang , Mohammad Zulkernine , and Anwar Haque

[6] William Stallings, "Cryptography and Network Security Principles and Practices", Third Edition, Prentice Hall, 2003.

[7] Java 2: The Complete Reference, Patrick Naughton and Herbert Schildt, Tata McGraw Hill, 1999.

[8] The Java Language Specification, 2nd ed, JamesGosling, Bill Joy, Guy Steele & Gilad Bracha, Sun Microsystems, 2000.

[9] ISS X-Force - www.iss.net/threats/ThreatList.php

[10] CERT - Carnegie Mellon University's Computer Emergency Response Team. www.cert.org/

[11] MD5 - Wikipedia, the free encyclopedia  http://en.wikipedia.org/wiki/MD5

**Authors**

**D.Kesavaraja** has completed his B.E Degree from the Department of Computer Science And Engineering from, Jayaraj Annapackiam CSI College of Engineering, Nazareth, Under Anna University, and Chennai in 2005. He  is currently pursuing the Masters in computer science and engineering at Manonmaniam Sundaranar University , Tirunelveli .He is a co-author of  a book titled "Fundamentals of Computing and Programming" ,ISBN 978-81-8472-099-0.   He is currently working as a Lecturer at the Department of Computer Science And Engineering, in Dr Sivanthi Aditanar College of Engineering. His research interests include Intrusion Detection, Web Development and Cloud Computing.

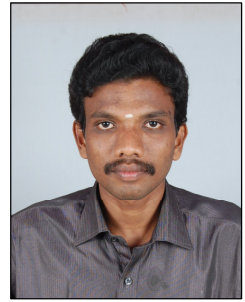

**R.Balasubramanian** received his B.E [Hons] degree in Computer Science and Engineering, from Bharathidhasan University in the year 1989. He completed his M.E, degree in Computer Science and Engineering, from Regional Engineering College Tiruchy in Bharathidhasan University in the year 1992.He is currently an Associate Professor in the department of Computer Science and Engineering at Manonmaniam Sundaranar University, Tirunelveli. He has a total of 20 years of Teaching Experience, out of which 18 years lie into research in the field of Digital Image Processing and Computing. Currently, he is pursuing his Doctorate from Manonmaniam Sundaranar University; Tirunelveli in the field of Digital Image Processing .He has published papers in 20 National Level Journals and many National and International Level Conferences. He has also coordinated and guided many research and PG Projects in the University .He is a co-author of the book titled "Computer Networks ", ISBN #812591238X .

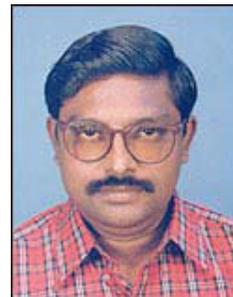

**D.Sasireka** has completed her B.Tech Degree from the Department of Information Technology from, Dr Sivanthi Aditanar College of Engineering, Tiruchendur , Under Anna University, and Chennai in 2008. She  is currently pursuing the Masters in Information Technology at Manonmaniam Sundaranar University , Tirunelveli .She is currently working as a Lecturer at the Department of Information Technology  in PSN College of Engineering Technology Melathedioor , Tirunelveli.. Her research interests include Intrusion Detection and Network Security  .

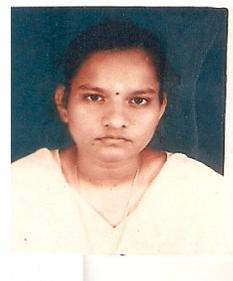